\documentstyle[12pt,a4]{article}
\normalsize
\def\simless{\mathbin{\lower 3pt\hbox{$\rlap{\raise 5pt\hbox{$\char'074$}}
\mathchar"7218$}}}
\def\simgreat{\mathbin{\lower 3pt\hbox{$\rlap{\raise 5pt \hbox{$\char'076$}}
\mathchar"7218$}}}
\catcode`@=11

\def\beqra{\begin{eqnarray}} \def\eeqra{\end{eqnarray}}
\def\beq{\begin{equation}}      \def\eeq{\end{equation}}

\def\fo{\hbox{{1}\kern-.25em\hbox{l}}}
\def\fnote#1#2{\begingroup\def\thefootnote{#1}\footnote{#2}\addtocounter
{footnote}{-1}\endgroup}

\def\ul{\underline}
\def\indt{\parindent2.5em}
\def\nd{\noindent}
\def\ch{\@startsection{section}{1}{\z@}{-3ex plus-1ex minus-.2ex}%
        {2ex plus.2ex}{\large\sc}}

\def\; \lapp \;{\raisebox{-.4ex}{\rlap{$\sim$}} \raisebox{.4ex}{$<$}}
\def\gapp{\raisebox{-.4ex}{\rlap{$\sim$}} \raisebox{.4ex}{$>$}}
\def\dt#1{{\buildrel{\hbox{\Large.}}\over{#1}}} % dot-over for sp/sb
\def\con{\ifmmode \hbox{\bf*} \else{\bf*}\fi}   % conjugation
\def\scon{\ifmmode \hbox{\footnotesize\rm\bf*} \else{\footnotesize\rm\bf*}\fi}
\def\bra#1{\left\langle #1\right|}      \def\ket#1{\left| #1\right\rangle}
\def\V#1{\langle#1\rangle}              \def\A#1{\left|#1\right|}
\def\sp#1{{}^{#1}}                      \def\sb#1{{}_{#1}}
\def\0#1{\relax\ifmmode\mathaccent"7017{#1}%    % puts a little circle atop,
        \else\accent23#1\relax\fi}              % as a halo of a saint

\def\inv#1{\frc{1}{#1}}                         % Like {1\over{#1}}
\def\ltsim{\matrix{<\cr\noalign{\vskip-7pt}\sim\cr}}
\def\gtsim{\matrix{>\cr\noalign{\vskip-7pt}\sim\cr}}
\def\haf{\frac{1}{2}}
\def\eslash{\not{\hbox{\kern-2pt $E$}}}

\catcode`@=12

\begin{document}
\hoffset=0.4cm
\voffset=-1truecm
\normalsize
\pagestyle{empty}
\def\ni{{\bar {N_i}}}    \def\nj{{\bar {N_j}}}   \def\n3{{\bar {N_3}}}
\def\li{\lambda_i}    \def\lj{\lambda_j}   \def\l3{\lambda_3}
\def\hn{h^\nu}       \def\hnij{h^\nu_{ij}}

\baselineskip=5pt
\begin{flushright}
SISSA-5/93/A
\end{flushright}
\begin{flushright}
DFPD/93/TH/03
\end{flushright}
\vspace{24pt}
\centerline{{\Large \bf The Singlet Majoron Model }}
\vskip 0.3 cm
\centerline{{\Large \bf with Hidden Scale Invariance}}
\vskip 1 cm
\begin{center}
{\large \bf Antonio Riotto}
\end{center}
\vskip 0.6 cm
\baselineskip=14pt
\centerline{\it
 International School for Advanced Studies,
SISSA,}
\centerline{\it via Beirut 2-4, I-34014 Trieste, Italy.}
\vskip 0.3 cm
\centerline{\it and}
\vskip 0.3 cm
\centerline{\it Istituto Nazionale di Fisica Nucleare,}
\centerline{\it Sezione di Padova, 35100 Padova, Italy}
\vspace{46pt}
\centerline{\large{\bf Abstract}}
\vskip 0.25 cm
\baselineskip=24pt
$~~~~$
We investigate an extension of the Singlet Majoron Model in which the
breaking of dilatation symmetry by the mass parameters of the scalar
potential is removed by means of a dilaton field. Starting from the
one-loop renormalization group improved potential, we discuss the ground
state of the theory. The flat direction in the classical potential is
lifted by quantum corrections and the true vacua are found. Studying the
finite temperature potential, we analyze the cosmological consequences of
a Jordan-Brans-Dicke dilaton and show that the lepton number is
spontaneously broken after the electroweak phase transition, thus
avoiding any constraint coming from the requirement of the preservation
of the baryon asymmetry in the early Universe. We also find that,
contrary to the Standard Model case, the dilaton cosmology does not
impose any upper bound on the scale of the spontaneous breaking of scale
invariance.
\pagebreak
%%%%%%%%%%%%%%%%%%%%%%%%%%%%%%%%%%%%%%%%%%%%%%%%%%%%%%%%%%%%%%%%%%%%%%%%%%
\normalsize
\def\simless{\mathbin{\lower 3pt\hbox{$\rlap{\raise 5pt\hbox{$\char'074$}}
\mathchar"7218$}}}
\def\simgreat{\mathbin{\lower 3pt\hbox{$\rlap{\raise 5pt \hbox{$\char'076$}}
\mathchar"7218$}}}
%%%%%%%%%%%%%%%%%%%%%%%%%%%%%%%%%%%%%%%%%%%%%%%%%%%%%%%%%%%%%%%%%%%%%%%%%%%
\catcode`@=11

\def\beqra{\begin{eqnarray}} \def\eeqra{\end{eqnarray}}
\def\beq{\begin{equation}}      \def\eeq{\end{equation}}

        %title page
\def\fo{\hbox{{1}\kern-.25em\hbox{l}}}
\def\fnote#1#2{\begingroup\def\thefootnote{#1}\footnote{#2}\addtocounter
{footnote}{-1}\endgroup}

        % formatting
\def\ul{\underline}
\def\indt{\parindent2.5em}
\def\nd{\noindent}
\def\ch{\@startsection{section}{1}{\z@}{-3ex plus-1ex minus-.2ex}%
        {2ex plus.2ex}{\large\sc}}

        % Math-related stuff
\def\; \lapp \;{\raisebox{-.4ex}{\rlap{$\sim$}} \raisebox{.4ex}{$<$}}
\def\gapp{\raisebox{-.4ex}{\rlap{$\sim$}} \raisebox{.4ex}{$>$}}
\def\dt#1{{\buildrel{\hbox{\Large.}}\over{#1}}} % dot-over for sp/sb
\def\con{\ifmmode \hbox{\bf*} \else{\bf*}\fi}   % conjugation
\def\scon{\ifmmode \hbox{\footnotesize\rm\bf*} \else{\footnotesize\rm\bf*}\fi}
\def\bra#1{\left\langle #1\right|}      \def\ket#1{\left| #1\right\rangle}
\def\V#1{\langle#1\rangle}              \def\A#1{\left|#1\right|}
\def\sp#1{{}^{#1}}                      \def\sb#1{{}_{#1}}
\def\0#1{\relax\ifmmode\mathaccent"7017{#1}%    % puts a little circle atop,
        \else\accent23#1\relax\fi}              % as a halo of a saint

\def\inv#1{\frc{1}{#1}}                         % Like {1\over{#1}}
\def\ltsim{\matrix{<\cr\noalign{\vskip-7pt}\sim\cr}}
\def\gtsim{\matrix{>\cr\noalign{\vskip-7pt}\sim\cr}}
\def\haf{\frac{1}{2}}
\def\eslash{\not{\hbox{\kern-2pt $E$}}}

\catcode`@=12

\hoffset=0.4cm
\voffset=-1truecm
\normalsize
\baselineskip=24pt
\setcounter{page}{1}
\def\ni{{\bar {N_i}}}    \def\nj{{\bar {N_j}}}   \def\n3{{\bar {N_3}}}
\def\li{\lambda_i}    \def\lj{\lambda_j}   \def\l3{\lambda_3}
\def\hn{h^\nu}       \def\hnij{h^\nu_{ij}}
\def\p{\phi} \def\m{m_{\phi}^{2}} \def\ms{m_{S}^{2}}
\def\mb{\bar{m}_{\phi}^{2}} \def\mbb{\bar{m}_{S}^{2}}
\def\s{\sigma}
\def\l{\lambda} \def\g{\gamma} \def\b{\beta}
\def\zp{z_{\phi}} \def\zs{z_{S}}
\def\gi{g_{R_{i}}}
\def\ap{\alpha_{\phi}}
\def\as{\alpha_{S}}
\def\v{\varepsilon\left(T\right)}
\def\a{\bar{z}_{\phi}^{1/2}\left(T\right)}
\voffset=-1 truecm
\normalsize
\begin{center}
{{\large {\bf 1. Introduction}}}
\end{center}
\vspace{0.5 truecm}

One of the most remarkable properties of the Standard Model (SM) of
electroweak interactions is that it is scale invariant up to the mass
term in the Higgs sector, responsible for the spontaneous symmetry
breaking, and a possible constant term related to the vacuum energy
density.

Theories with mass parameters may still have {\it hidden} nonlinearly
realized scale invariance \cite{Coleman}, which may play a role in
solving the cosmological constant problem \cite{Peccei} and can account
for the hierarchy of scales observed in the SM between the weak scale
and the Planck and/or the Grand Unification scale \cite{hie}. Also it is
intriguing that models which display a nonlinearly realized scale
invariance in their low energy effective theories occur in some
unification schemes, including models based on compactification of
higher dimensions \cite{Fre} and string theories \cite{Witt}.

A trace of the scale invariance may still exist at low energies in the
form of a pseudo-Goldstone boson, the dilaton, which couples in a
universal way to all mass terms and gains a mass due to the explicit
breaking of classical scale invariance by quantum corrections
\cite{hie}. Coupling a Jordan-Brans-Dicke (JBD) dilaton to the SM leads
to a model equivalent under a conformal transformation to the SM coupled
to a JBD gravity theory \cite{Jor} and to interesting phenomenological
consequences. In particular, Buchm\"uller and Busch \cite{hie} have
shown that the existence of a JBD dilaton would impose an upper bound of
about 100 GeV on the top quark mass, the precise value depending on the
dilaton decay constant $f$,  from the requirement that the ground state
breaks the electroweak gauge symmetry.

The presence of a JBD dilaton would also have significant cosmological
implications, for the range of cosmologically allowed neutrino masses
\cite{neut} and for the dynamics of the electroweak phase transition,
see refs. \cite{phase} and \cite{Mc}, which has been shown to be first
order and to occur at the chiral symmetry breaking phase transition. In
particular, Mc Donald \cite{Mc} has recently emphasized that there is a
serious problem with the energy density in the dilaton field following
the electroweak phase transition, with most of the energy in the
electroweak vacuum going into oscillations of the dilaton field at
$T_{QCD}\sim$ 150 MeV, causing the Universe to be matter-dominated at
nucleosynthesis and the standard calculation of element abundances to
disagree with observations. To avoid this problem one has to impose the
upper bound $f\simless 10^{7}$ GeV on the scale of the spontaneous
breaking of scale invariance. Such a requirement is well in contrast
with the natural identification $f\simeq M_{P}$ in models in which the
hierarchy between the Planck scale $M_{P}$ and the weak scale is
explained by the underlying scale invariance \cite{hie} through the
introduction of a JBD dilaton. Motivated by this troublesome
inconsistency present in the SM, we investigate which kind of
implications would have have the coupling of a JBD dilaton to one of the most
attractive extension of the SM, the Singlet Majoron Model (SMM)
\cite{smm}. In the SMM a gauge singlet scalar and three right-handed
neutrinos are introduced, and a global $U(1)_L$ group
associated to the lepton number is spontaneously broken, giving rise to
Majorana masses for the right-handed neutrinos. The model naturally
incorporates the see-saw mechanism \cite{ram} and can therefore explain
why left-handed neutrinos are much lighter than their right-handed
counterparts. However, the realization that the baryon ($B$) and lepton
$(L)$ violating, while $(B-L)$ conserving, quantum effects in the SM are
efficient at high temperatures \cite{sph} gives rise to a very strong
bound on the Majorana masses of light neutrinos, $m_{\nu}\simless$ 1 eV
\cite{har}. The point is that the SM anomalous
effects still allow a baryon asymmetry generated at some
superheavy scale \cite{super} to survive if the Universe had a
nonvanishing primordial $(B-L)$ asymmetry, but, if other interactions which
violate $B$, $L$ and also the combination $(B-L)$ are in
equilibrium at temperatures above the Fermi scale, then no cosmological
baryon asymmetry can survive \cite{har}. One has to require that the new
$L$-violating interactions not to be in equilibrium at all temperatures
at which anomalous interactions are still active, which leads to the
above mentioned strong bound on $m_{\nu}$. A natural way to avoid such a
constraint is to generate Majorana neutrino masses with a spontaneous
$L$-number breaking at the electroweak scale or below. The conservation
of $L$-number at higher scales prevents the existence of the dangerous
$\Delta
L=2$ effective operator $\left(m_{\nu}/\left\langle \phi
\right\rangle^2\right)\left(L_{L}\phi\right)^2$, where $\phi$ is the
standard Higgs scalar doublet and $L_{L}$ is a lepton doublet,
$L_{L}=\left(\nu_{L},l_{L}\right)$. It has been recently shown that in
the supersymmetrized version of the SMM \cite{riotto} and in the
triplet-singlet Majoron Model \cite{gel} the phase transition leading to
the $L$-number breaking can occur at temperatures below the weak scale,
thus avoiding any constraint coming from the requirement of the
preservation of the baryon asymmetry.

 The aim of this paper is to show
that the same feature is naturally achieved when a JBD dilaton is
coupled to the SMM and that, contrary to what happens in the SM,
 no upper bound on the dilaton decay
constant $f$ comes from considerations about the JBD dilaton cosmology.

The paper is organized as follows. In Sect. 2 we describe the SMM with
hidden scale invariance and derive the one-loop renormalization group
improved effective potential, whose minimization allows to break the
vacuum degeneracy present at the classical level. In Sect. 3 we deal
with the one-loop finite temperature effective potential and study the
dynamics of the $SU(2)_{L}\otimes U(1)_{Y}$ and $U(1)_{L}$ phase
transitions. In Sect. 4 we present our conclusions.
\newpage
\begin{center}
{{\large {\bf 2. The SMM with hidden scale invariance}}}
\end{center}
\vspace{0.5 truecm}

The Lagrangian for the scale invariant extension of the SMM is
\beqra
{\cal L}&=& \frac{1}{2}{\rm
e}^{2\s/f}\partial_{\mu}\s\partial^{\mu}\s +
\left(D_{\mu}\p\right)^{\dag}\left(D^{\mu}\p\right) +
\left(\partial_{\mu}S\right)^{\dag}\left(\partial^{\mu}S\right) +
i \bar{N}_{R}\not\!\partial N_{R}\nonumber\\
&-&\left\{h_{\nu}\bar{L}_{L}\p N_{R} + \frac{1}{2}g_{R}
\bar{N}_{R} N_{R}^{c} S +
{\rm h.c.}\right\} - V_{0}\left(\s,\p,S\right),
\eeqra
where
\beqra
V_{0}\left(\s,\p,S\right)&=& \bar{a}^{4} + \mb \left(\p^{\dag}\p\right) +
\mbb \left(S^{\dag}S\right)\nonumber\\
&+&
\gamma\left(\p^{\dag}\p\right)\left(S^{\dag}S\right) +
\frac{\l}{2}\left(\p^{\dag}\p\right)^2 +
\frac{\b}{2}\left(S^{\dag}S\right)^2
\eeqra
and
\[
\bar{a}^{4}\equiv a^4~{\rm e}^{4\s/f},~~~~~~ \mb\equiv \m~{\rm
e}^{2\s/f},~~~~~~ \mbb\equiv\ms~{\rm
e}^{2\s/f}, ~~~~~~\l>0,~~~~~~\b>0.
\]
Here $\s$ is the dilaton field, $f$ its decay constant, $\p$ is the
scalar Higgs doublet, $D_{\mu}$ is the $SU(2)_{L}\otimes U(1)_{Y}$ gauge
covariant derivative, $S$ is the gauge singlet field carrying lepton
number $L=2$, $N_{R}$ is the gauge singlet right-handed neutrino
$\left(N^{c}\equiv C \bar{N}^{T}\right)$, $L_{L}$ is the lepton doublet
and $h_{\nu}$ and $g_{R}$ are $3\times 3$ matrices of Yukawa couplings.

Due to the specific couplings of the Goldstone field $\s$, the
Lagrangian (1) is invariant under dilatations
\beqra
\delta\s=\delta\alpha\left(f+x^{\mu}\partial_{\mu}\s\right)~~~~&,&~~~~
\delta\p=\delta\alpha\left(\p+x^{\mu}\partial_{\mu}\p\right),\nonumber\\
\delta S=\delta\alpha\left(S+x^{\mu}\partial_{\mu}S\right)~~~~&,&~~~~
\delta \psi_{L\left(R\right)R}=\delta\alpha
\left(\frac{3}{2}\psi_{L\left(R\right)}+x^{\mu}\partial_{\mu}
\psi_{L\left(R\right)}\right),
\eeqra
where $\psi$ denotes a generical fermion field in the Lagrangian (1).
The classical equations of motion for the scalar fields read
\beqra
&& D_{\mu}D^{\mu}\p + \mb\p + \l\left(\p^{\dag}\p\right)\p + \g
\left(S^{\dag}S\right)\p=0,\nonumber\\
&& \partial_{\mu}\partial^{\mu} S + \mbb S + \b \left(S^{\dag}S\right) +
\g\left(\p^{\dag}\phi\right)S=0,\\
&&{\rm e}^{2\s/f} \left(\partial_{\mu}\partial^{\mu}\s + \frac{2}{f}
\partial_{\mu}\s\partial^{\mu}\s\right) + \frac{2}{f}\mb
\left(\p^{\dag}\p\right) + \frac{2}{f}\mbb\left(S^{\dag}S\right) +
\frac{4}{f}\bar{a}^{4}=0.\nonumber
\eeqra
The existence of non-trivial constant solutions $\s_{0}$, $\p_{0}$ and
$S_{0}$ constraints the allowed parameters present in $V_{0}$. For
instance, for $\m>0$, $\ms>0$ and $\g>0$, the only stationary points
are $\p_{0}=0$, $S_{0}=0$ and $\s_{0}$ remains undetermined. In the
following we shall choose $\m$ and $\ms$ both negative. In such a case
from eq. (4) one discovers that the constant $a^4$ is fixed to be
\beq
a^4=-\frac{1}{2}\frac{2\g\m\ms-\b m_{\p}^{4}-\l m_{S}^{4}}{
\b\l-\g^2}
\eeq
and symmetry breaking vacuum expectation values define in the valley floor
a flat direction
\begin{eqnarray}
\p^{\dag}_{0}\p_{0}&\equiv& v^{2} = \frac{\g\mbb -
\b\mb}{\b\l-\g^{2}}>0,
\nonumber \\
S_{0}^{\dag}S_{0}&\equiv& \bar{f}^{2} =
\frac{\g\mb-\l\mbb}{\b\l-\g^{2}}>0,\\
{\rm e}^{2\s_{0}/f}&=&-\frac{\b S^{\dag}_{0}S_{0}+\g
\p^{\dag}_{0}\p_{0}}{\ms}=-\frac{\l \p^{\dag}_{0}\p_{0} + \g
S_{0}^{\dag}S_{0}}{\m},\nonumber
\end{eqnarray}
with
\[
\g^{2}<\b\l.
\]
Note that $\g>0$ should be bounded from above in order that symmetries
are broken. (If $\g<0$ the same condition is required for the potential
to be bounded from below).

The consistency requirements for the couplings present in $V_{0}$ imply
that the classical energy density vanishes at the stationary points
and the potential $V_{0}$ takes the special form
\beqra
V_{0}\left(\s,\p,S\right)&=&\frac{\l}{2}\left[\p^{\dag}\p-\frac{
{\rm e}^{2\s/f}}{\b\l-\g^{2}}\left(\g\ms-\b\m\right)\right]^2\nonumber\\
&+& \frac{\b}{2}\left[S^{\dag}S-\frac{{\rm e}^{2\s/f}}{\b\l-\g^2}\left(
\g\m-\l\ms\right)\right]^2\nonumber\\
&+&\g \left[\p^{\dag}\p-\frac{
{\rm e}^{2\s/f}}{\b\l-\g^{2}}\left(\g\ms-\b\m\right)\right]\nonumber\\
&\times&
\left[S^{\dag}S-\frac{{\rm e}^{2\s/f}}{\b\l-\g^2}\left(
\g\m-\l\ms\right)\right].
\eeqra
The degeneracy in the vacuum expectation values and the vanishing of the
vacuum energy density are expected to disappear in quantum field
theory where scale invariance is anomalous \cite{an}. Thus, we consider
the one-loop corrections to the potential $V_{0}$ which can be computed
by standard methods \cite{we}. Since the dilaton interactions are not
manifestly renormalizable, we treat $\s$ as a classical background
field. A convenient choice of the renormalization conditions yields, in
the Landau gauge, $\left(\zp\equiv \p^{\dag}\p,
\zs\equiv S^{\dag}S\right)$
\beqra
V\left(\s,\zp,\zs\right)&=& \bar{a}^{4} + \mb \zp +
\mbb \zs + \gamma \zp\zs + \frac{\l}{2}\zp^2 +
\frac{\b}{2}\zs^2\nonumber\\
&+&\frac{1}{(8\pi)^2}\left\{\frac{3}{2}\left(\mb+\l\zp+\g\zs\right)^2\left[
{\rm ln}\frac{\left(\mb+\l\zp+\g\zs\right)^2}{M^4}-1\right]\right.\nonumber\\
&+&\frac{1}{2}\left(\mbb+\l\zp+\g\zs\right)^2\left[{\rm ln}\frac{
\left(\mbb+\l\zp+\g\zs\right)^2}{M^4}-1\right]\nonumber \\
&+&\frac{1}{2}m_{+}^4\left[{\rm
ln}\left(\frac{m_{+}^{4}}{M^4}\right)-1\right] +
\frac{1}{2}m_{-}^4\left[{\rm
ln}\left(\frac{m_{-}^{4}}{M^4}\right)-1\right] \nonumber \\
&+&\frac{3}{2} g^4\zp^2\left[{\rm
ln}\left(\frac{g^2\zp}{2M^2}\right)-\frac{1}{2}\right]+\frac{3}{4}
\left(g^2+{g^{\prime}}^2\right)^2\zp^2\left[{\rm
%% FOLLOWING LINE CANNOT BE BROKEN BEFORE 80 CHAR
ln}\frac{\left(g^2+{g^{\prime}}^2\right)\zp}{2M^2}-\frac{1}{2}\right]\nonumber\\
&-&12g_{t}\zp^2 \left.\left[{\rm
ln}\left(\frac{g_{t}^2\zp}{M^2}\right)-\frac{1}{2}\right]-2\sum_{i=1}^{3}
\gi\zs^2\left[{\rm
ln}\left(\frac{\gi^{2}\zs}{M^2}\right)-\frac{1}{2}\right]\right\},
\eeqra
where $g$ and $g^{\prime}$ are the $SU(2)_{L}$ and the $U(1)_{Y}$ gauge
couplings, respectively, $g_{t}$ is the Yukawa coupling for the top
quark, the heaviest fermion in the SM and $m_{\pm}^2$ are given by
\beqra
m_{\pm}^2&=&\frac{1}{2}\left[\mb+\mbb+3\l\zp+3\b\zs+\g\left(\zp+\zs\right)
\right.
\nonumber\\ &\pm&
\left.\sqrt{\left(\mb+3\l\zp-\mbb-3\b\zs+\g\zp-\g\zs\right)^2+
16\g^2\zp\zs}\right].
\eeqra
The parameters $a^4$, $\m$, $\ms$, $\l$, $\b$ and $\g$ are now
renormalized parameters which depend on the renormalization mass $M$.
Their $M$-dependence can be read off from eq. (8)
\beqra
M\frac{da^4}{dM}&=&\frac{1}{16\pi^2}\left(2
m_{\p}^4+m_{S}^4\right),\\
M\frac{d \ms}{dM}&=&\frac{1}{16\pi^2}\left(4\b\ms +\g\ms +3\g\m\right)+
2\gamma_{S}\ms,\\
M\frac{d
\m}{dM}&=&\frac{1}{16\pi^2}\left(6\l\m+\g\ms+\g\m\right)+2\gamma_{\p}\m,\\
M\frac{d
\l}{dM}&=&\frac{3}{64\pi^2}\left[\left(g^2+{g^{\prime}}^2\right)^2
+2g^4-16g_{t}^4 +16\l^2+\frac{8}{3}\l\g\right]+4\g\g_{\p},\\
M\frac{d \b}{dM}&=&\frac{1}{32\pi^2}\left(8\g^2 + 20\b^2 - \sum_{i=1}^3
\gi^4 + 12\b\g\right)+ 4\g_{S}\b,\\
M\frac{d \g}{dM}&=&\frac{1}{16\pi^2}\left(3\l+\b+4\g\right)\g + 2\left(
\g_{\phi}+\g_{S}\right)\g,
\eeqra
where $\g_{\p}$ and $\g_{S}$ are the anomalous dimensions of the Higgs
and singlet fields
\beqra
\g_{\phi}&=&\frac{1}{64\pi^2}\left(-9g^2-3{g^{\prime}}^2+12
g_{t}^2\right),\\
\g_{S}&=& \frac{1}{32\pi^2}\sum_{i=1}^3\gi^2.
\eeqra
The constraint given by eq. (5) can be also imposed on the renormalized
quantities and, since the scale dependence of
$\left(\b\l-\g^2\right)a^4$ and
$\left(2\g\m\ms-\b m_{\p}^4-\l m_{S}^4\right)$ is different, the
renormalization mass $M$ remains fixed.

The minimization of the one-loop renormalization group effective potential
allows us to eliminate the vacuum degeneracy described by eq. (6). The
analysis may be simplified considerably by expanding the complicated
terms in eq. (8) in a series around $\g=0$ and retaining only the lowest
order terms. A straightforward calculation gives
\beq
\left(\p^{\dag}\p\right)_{0}=\frac{1}{2\l}A_{\p}\left[
1-\frac{\l}{16\pi^2}{\rm ln}\frac{A_{\p}}{M^2}\right],~~~~~
\left(S^{\dag}S\right)_{0}=\frac{1}{2\b}A_{S}\left[
1-\frac{\b}{16\pi^2}{\rm ln}\frac{A_{S}}{M^2}\right],
\eeq
where
\beqra
A_{\p}&=&{\rm
exp}\left[\frac{A_{1}^{\p}-A_{2}^{\p}}{B}\right]M^2,\nonumber\\
A_{S}&=&{\rm
exp}\left[\frac{A_{1}^{S}-A_{2}^{S}}{B}\right]M^2,\nonumber\\
A_{1}^{\p}&=&K_{\p}K_{S}\left[-\frac{3}{2\l}g^4{\rm ln}\frac{g^2}{4\l}
-\frac{3}{4\l}\left(g^2+ {g^{\prime}}^2\right)^2{\rm
ln}\frac{g^2+{g^{\prime}}^2}{4\l}+\frac{12}{\l}g_{t}^4{\rm
ln}\frac{g_{t}^2}{2\l}\right]\nonumber\\
&\times&\left(4\b-\sum_{i=1}^3\frac{\gi^4}{2\b}\right),\nonumber\\
A_{2}^{\p}&=&8\g K_{S}^2\sum_{i=1}^3\frac{\gi^4}{2\b}{\rm
ln}\frac{\gi^2}{2\b},\nonumber\\
A_{1}^{S}&=&4K_{\p}K_{S}\sum_{i=1}^3\frac{\gi^4}{2\b}{\rm
ln}\frac{\gi^2}{2\b}\left[4\l + \frac{3}{2\l}g^4 + \frac{3}{4\l}\left(g^2+
{g^{\prime}}^2\right)^2-\frac{12}{\l}g_{t}^4\right],\nonumber\\
A_{2}^{S}&=&2\g K_{\p}^2\left[-\frac{3}{2\l}g^4{\rm ln}\frac{g^2}{4\l}
-\frac{3}{4\l}\left(g^2+ {g^{\prime}}^2\right)^2{\rm
ln}\frac{g^2+{g^{\prime}}^2}{4\l}+\frac{12}{\l}g_{t}^4{\rm
ln}\frac{g_{t}^2}{2\l}\right],\nonumber\\
B&=&K_{\p}K_{S}\left[\left(4\l + \frac{3}{2\l}g^4 + \frac{3}{4\l}\left(g^2+
{g^{\prime}}^2\right)^2-\frac{12}{\l}g_{t}^4\right)\right.\nonumber\\
&\times&
\left.\left(4\b-\sum_{i=1}^3\frac{\gi^4}{2\b}\right)-4\g^2\right],
\eeqra
and
\beqra
K_{\p}&=&{\rm exp}\left[-\frac{3g^4{\rm ln}\frac{g^2}{4\l}+\frac{3}{2}
\left(g^2+{g^{\prime}}^2\right)^2{\rm ln}\frac{g^2+{g^{\prime}}^2}{4\l}
-24g_{t}^4{\rm ln}\frac{g_{t}^2}{2\l}}{8\l^2 + 3g^4 + \frac{3}{2}
\left(g^2+{g^{\prime}}^2\right)^2-24g_{t}^4}\right],\nonumber\\
K_{S}&=&{\rm exp} \left[\frac{4\sum_{i=1}^3\gi^4{\rm
ln}\frac{\gi^2}{2\b}}{8\b^2-4\sum_{i=1}^3\gi^4}\right].
\eeqra
The value of ${\rm exp}\left(2\s_{0}/f\right)$ can be read off from eqs.
(6) and (18)-(20) where all the parameters are meant to be evaluated at
the scale $M$.

To find the ground-state energy and the dilaton mass, it is usuful to
define the new fields \cite{Coleman}
\beq
\tilde{\p}\equiv {\rm e}^{-\s/f}~\p,~~~~~~\tilde{S}\equiv
{\rm e}^{-\s/f}~ S,
\eeq
so that the one-loop effective potential (8) now reads
\beq
V\left(\s,\tilde{\p},\tilde{S}\right)={\rm e}^{4\s/f}~\left[
\bar{V}\left(\tilde{\p},\tilde{S}\right)+ \Delta \left(\tilde{\p},\tilde{S}
\right)\frac{\s}{f}\right],
\eeq
where $\bar{V}$ is the one-loop effective potential without dilaton
field and $\Delta$ is the anomalous divergence of the dilatation current
\beq
\Delta=-M\frac{\partial}{\partial
M}\bar{V}\left(\tilde{\p},\tilde{S}\right).
\eeq
Using the extremum conditions
\[
\left.\frac{\partial V}{\partial \tilde{\p}}\right|_{\tilde{\p}_{0},
\tilde{S}_{0},\s_{0}}=0,~~~~
\left.\frac{\partial V}{\partial \tilde{S}}\right|_{\tilde{\p}_{0},
\tilde{S}_{0},\s_{0}}=0,~~~~
\left.\frac{\partial V}{\partial \s}\right|_{\tilde{\p}_{0},
\tilde{S}_{0},\s_{0}}=0,
\]
and the fact that $\bar{V}$ is scale-independent,
$\left(d\bar{V}/dM\right)=0$, one easily derives the ground-state energy
and the dilaton mass \cite{hie}
\beqra
\left\langle V\right\rangle_{0}&=&\frac{1}{4}\left\langle M\frac{
\partial\bar{V}}{\partial M}\right\rangle_{0}=-\frac{1}{128\pi^2}
{\rm Str}{\cal M}^4,\\
m_{\s}^2&=&-\frac{4}{f^2}\frac{1}{4}\left\langle M\frac{
\partial\bar{V}}{\partial M}\right\rangle_{0}=\frac{
1}{8\pi^2 f^2}{\rm Str}{\cal M}^4,\\
{\rm Str}{\cal M}^4&=& 6m_{W}^4+3m_{Z}^4+m_{+}^4+m_{-}^4-12 m_{t}^4-
2\sum_{i=1}^3m_{N_{R_{i}}}^4.
\eeqra
In ref. \cite{neut} the upper bound $m_{N_{R}}\simless 130$ GeV on the
largest right-handed neutrino mass was obtained, roughly requiring that
$m_{\s}^2>0$ or, equivalently, $m_{t}^4+(1/6)\sum_{i=1}^3
m_{N_{R_{i}}}^4\simless\left(100\, \mbox{GeV}\right)^4$. However, in ref.
\cite{neut} it was assumed that neutrino masses are present and the
lepton number is broken at all scales. In the model under consideration
the presence of the gauge singlet $S$ is fundamental to provide neutrino
masses via the spontaneous breaking of the global group
$U(1)_{L}$ at the scale $\bar{f}$ and one has to require that
$m_{t}^4+(1/6)\sum_{i=1}^3
m_{N_{R_{i}}}^4-(1/12)m_{-}^4\simless\left(100\, \mbox{GeV}\right)^4$,
which does not give any particularly strong bound on the largest eigenvalue
$m_{N_{R}}$. Nevertheless, in the see-saw mechanism one would obtain for
the electron neutrino mass $m_{\nu_{e}}\sim
m_{D}^2/\left(g_{R}\bar{f}\right)$, where in the simplest scenario one
expects $m_{\nu_{e}}\sim m_{e}$. The current limit on $m_{\nu_{e}}$
would then imply that $\bar{f}\simgreat$ 100 GeV. If $\bar{f}\gg v$,
then the $\p$- and the $S$-sectors would be effectively decoupled unless
one imposes a fine-tuning on the involved parameters or, more correctly,
a fine-tuning on the initial conditions for the Renormalization Group
Equations (10)-(15). Thus, in the following we shall assume that
$\bar{f}$ and $v$ are roughly the same scale, as well as $\m$ and $\ms$.

We can embed our theory in a
background spacetime described by a metric $g_{\mu\nu}$ and demand that
the resulting theory be invariant under local rescalings of the metric
$g_{\mu\nu}\rightarrow g_{\mu\nu}^{\prime}={\rm
exp}\left[2\g(x)\right]g_{\mu\nu}$. In order that the potential term
$\sqrt{-g}V\left(\s,\p,S\right)$ be invariant, we demand that $\s$, $\p$
and $S$ transform as $\s^{\prime}=\s-f\g(x)$ and
$\p^{\prime}={\rm
exp}\left[-\g(x)\right]\p$ and $S^{\prime}=
{\rm exp}\left[-\g(x)\right]S$ \cite{in}. The $\p$ and the $S$
kinetic terms used in eq. (1) are not Weyl invariant. However,
if we use for the
terms
involving the dilaton field $\s$ and the Higgs doublet $\p$ (and
similarly
for $S$) the Lagrangian
\beqra
{\cal L}&=& \sqrt{-g}
\left[\frac{1}{2}g^{\mu\nu}\partial_{\mu}\s\partial_{\nu}\s -\frac{1}{f}
g^{\mu\nu}\partial_{\mu}\left(\p^{\dag}\p\right)\partial_{\nu}\s+\frac{1}{f^2}
g^{\mu\nu}\partial_{\mu}\s\partial_{\nu}\s
\left(\p^{\dag}\p\right)\right.\nonumber\\
&+&
\left. g^{\mu\nu}\left(D_{\mu}\p\right)^{\dag}\left(D_{\nu}\p\right)
-V\left(\s,\p\right)\right],
\eeqra
the classical theory is no longer invariant under dilatations, which
would require the kinetic term ${\rm exp}\left(2\s/f\right)
\partial_{\mu}\s\partial^{\mu}\s$ for the dilaton field. However, a
characteristic feature of the Lagrangian (27) is an approximate Weyl
invariance in curved space, which is only broken by the kinetic terms of
gravitational and dilaton field
\beqra
\sqrt{-g}\left(\frac{1}{2}\kappa{\cal R}+\frac{1}{2}
\partial_{\mu}\s\partial^{\mu}\s\right)&=&\sqrt{-\bar{g}}\left(
D\bar{{\cal R}}+\frac{\omega}{D}\bar{g}^{\mu\nu}\partial_{\mu}D\partial
_{\nu}D\right),\nonumber\\
\bar{g}_{\mu\nu}&=&{\rm e}^{2\s/f}~ g_{\mu\nu},
\eeqra
where ${\cal R}$ is the curvature scalar, $\omega=f^2/4\kappa$ and
$D=(1/2){\rm exp}\left[-2\s/\left(f^2+6\kappa\right)^{1/2}\right]$ is
what is generally called
the JBD dilaton. The theory is the Jordan-Brans-Dicke theory of gravity
with the SMM as matter sector \cite{Jor}.
In the next section we shall make use of kinetic terms
like those in eq. (27) to find the Friedmann-Robertson-Walker (FRW)
equation governing the dynamics of the dilaton field.
\newpage
\begin{center}
{{\large {\bf 3. The effect of dilatons on the phase transitions in SMM}}}
\end{center}
\vspace{0.5 truecm}

We first consider the finite temperature potential \cite{dolan} of the
SMM where the dilaton can be added in the standard manner \cite{Coleman}
\beqra
V_{T}\left(\s,\zp,\zs\right)&=& c\pi^2 T^4+ \bar{a}^{4} + \mb \zp +
\mbb \zs + \gamma \zp\zs + \frac{\l}{2}\zp^2 +
\frac{\b}{2}\zs^2\nonumber\\
&+& \frac{1}{6}\mb T^2 + \frac{1}{12}\mbb T^2 +\ap T^2\zp +\as
T^2\zs\nonumber\\
&-&\frac{1}{12\pi}\left[3\left(\mb+\ap T^2+\l\zp+\g\zs\right)^{3/2}
+\left(\mbb +\as T^2+\b\zs +\g\zp\right)^{3/2}\right.\nonumber\\
&+& \left.m_{+}^3 +m_{-}^{3} + 6 \left(\frac{g^2}{2}\right)^{3/2}\zp^{3/2}
+3\left(\frac{g^2+{g^{\prime}}^2}{2}\right)^{3/2}\zp^{3/2}\right]\nonumber\\
&-&\frac{1}{\left(8\pi\right)^2}\left\{\left[3\left(\mb+\ap T^2 +\l\zp+
\g\zs\right)^2 +\left(\mbb+\as T^2
+\b\zs+\g\zp\right)^2\right.\right.\nonumber\\
 &+& \left. m_{+}^4 + m_{-}^4+\frac{3}{2}g^4\zp^2+\frac{3}{4}
\left(g^2+{g^{\prime}}^2\right)^2\zp^2\right]\left[{\rm ln}\left(
\frac{M^2}{A_{b}T^2}\right)+\frac{1}{2}\right]\nonumber\\
&-&\left.\left[12 g_{t}^4\zp^2+2\sum_{i=1}^3\gi^4\zs^2\right]
\left[{\rm ln}\left(
\frac{M^2}{A_{f}T^2}\right)+\frac{1}{2}\right]\right\},
\eeqra
where
\beqra
\ap&=&\frac{1}{24}\left(6\l+2\g+\frac{9}{2}g^2+\frac{3}{2}
{g^{\prime}}^2+6g_{t}^2\right),\nonumber\\
\as&=&\frac{1}{24}\left(4\b +4\g +\sum_{i=1}^{3}\gi^2\right),\nonumber\\
A_{f}&=&\pi^2{\rm exp}\left(\frac{3}{2}-2\g_{E}\right),~~~ A_{b}=16~
A_{f},~~~\gamma_{E}\simeq 0.57.
\eeqra
The finite temperature potential $V_{T}$ is given in an expansion
of $(1/\pi)m_{i}/T$,
up to terms of order
$1/T$, where $m_{i}\left(\s,\zp,\zs\right)$ are the various particle masses
as functions of the scalar fields. The constant $c$ essentially counts
the total number of degrees of freedom. We have replaced the
bare mass parameters by the corresponding plasma masses \cite{dolan} in
$V_{T}$ to avoid the latter to be ill-defined in $\zp=\zs=0$.

The dilaton field $\s$ is not in thermal equilibrium at
temperatures below the scale $f$ (see Enqvist in ref. \cite{phase}).
Since $\left|\dot{\s}/\s\right|\ll T$, from the point of view of finite
temperature field theory the $\s$ field will act as a constant
\cite{Mc}. The dynamics of the dilaton field is described by the FRW
equation of motion
\beqra
\ddot{\s}+3H\dot{\s}+\frac{2}{f^2}\ddot{\s}\left(\zp+\zs\right)
&+&\frac{2}{f^2}\dot{\s}\left(\dot{\zp}+
\dot{\zs}\right)
-\frac{1}{f}\left(\ddot{\zp}+
\ddot{\zs}\right)\nonumber\\
&=&-\frac{2}{f}{\rm e}^{2\s/f}\left(2a^4{\rm e}^{2\s/f}+\m
\zp+\ms \zs\right),
\eeqra
where $H$ is the Hubble constant. In writing eq. (31) we have
followed ref. \cite{Mc} and taken into
account that $\s$ does not fluctuate fast enough to be in thermal
equilibrium with $\p$ and $S$, so that its expectation value should not be
determined by minimizing the finite temperature potential, but the $T=0$
potential where $\zp$ and $\zs$ have to be replaced in eq. (31) by their
thermal average values, $\langle\zp\rangle_{T}$ and
$\langle\zs\rangle_{T}$, respectively. The value of $\s$ at its minimum
at a given temperature is therefore
\beq
{\rm exp}\left(\frac{2\langle\s\rangle}{f}\right)=k_{\p}\langle\zp\rangle_{T}
+k_{S}\langle\zs\rangle_{T},
\eeq
where
\beq
k_{\p}\equiv \frac{\m\left(\g^2-\b\l\right)}{\l m_{S}^4+
\b m_{\p}^4-2\g\m\ms},~~~~~ k_{S}\equiv
\frac{\ms\left(\g^2-\b\l\right)}{\l m_{S}^4+
\b m_{\p}^4-2\g\m\ms}.
\eeq
In the range of validity
of the high temperature expansion
of $V_{T}$, we can approximate $\langle\zp\rangle_{T}=\langle\zs\rangle_{T}
\simeq T^2/12$ \cite{lin} and find that
\beqra
V_{T}&=&\tilde{c}\pi^4T^4 + a_{\p}T^2\zp+ a_{S}T^2\zs
+\frac{\l}{2}\zp^2+\frac{\b}{2}\zs^2 \nonumber\\
&+&\g \zp\zs +
{\cal O}\left(\zp^{3/2},\zs^{3/2}\right),
\eeqra
where
\beqra
a_{\p}&\equiv& \ap+\frac{\m}{12}\left(k_{\p}+k_{S}\right),\nonumber\\
a_{S}&\equiv& \as+\frac{\ms}{12}\left(k_{\p}+k_{S}\right),
\eeqra
and $\tilde{c}$ can be easily inferred from eqs. (29) and (32).

Since we are presently interested in the case
$\left|\dot{\s}/\s\right|\gg H$, we expect that the value of
$\left|\dot{\s}/\s\right|$ is set by the mass of $\s$ at the minimum
of its potential \cite{Mc}
\beq
m_{\s}^2=\left.\frac{\partial^2 V}{\partial \s^2}\right|_{\s=
\langle\s\rangle}\simeq \frac{\l}{f^2}\frac{T^4}{18},
\eeq
where we have assumed $\l\sim \b$ and $\m\sim\ms$. Thus
$\left|\dot{\s}/\s\right|$ will be much smaller than $T$ if
\beq
T^2\ll \frac{18 T^2}{\l},
\eeq
which is certainly satisfied for the range of temperatures under
consideration (one can also easily show that
the assumption $\left|\dot{\s}/\s\right|\gg H$ is well satisfied for
$T\simless f$).

Looking at eqs. (34) and (35), we discover
that, if $a_{\p}$ and $a_{S}$ are both positive, the $SU(2)_{L}\otimes
U(1)_{Y}$ and the $U(1)_{L}$ symmetries will be unbroken at any temperature.
Moreover, we cannot conclude that, if $a_{\p}$ and $a_{S}$ are
negative, the same symmetries are always broken since our finite
temperature expansion is only valid for $a_{\p},a_{S}>0$, so that no
firm conclusion concerning the symmetry breaking can be drawn in this
case. In the following we shall assume $a_{\p}$ and $a_{S}$ both
positive, which is likely to be true if $\lambda\sim\b$. Thus the finite
temperature corrections generate a barrier at $\zp=\zs=0$ which persists
down to $T=0$.

Since in the SMM with dilaton the vacuum energy density vanishes at the
tree
level, we assume that the critical temperature $T_{c}$, which is defined
by \cite{phase}
\beq
V_{T_{c}}(0)=V_{T_{c}}\left(v^{2}, \bar{f}^2\right),
\eeq
is smaller than $v\sim\bar{f}$. If in the broken phase particle masses
are either much smaller or much larger than the critical temperature,
the difference in the finite temperature correction to the effective
potential at $\left(\zp=0,\zs=0\right)$ and $\left(\zp=v^2,\zs=
\bar{f}^2\right)$ is simply determined by the numbers of effectively
massless particles in both phases \cite{ad} and $T_{c}$ results
\beq
T_{c}=\left\{\frac{45}{1712\pi^4}\left[
6m_{W}^4+3m_{Z}^4+m_{+}^4+m_{-}^4-12m_{t}^4-2\sum_{i=1}^3
m_{N_{R_{i}}}^4\right]\right\}^{1/4}={\cal O}\left(
10\,\mbox{GeV}\right).
\eeq

Using the numerical results found in ref. \cite{kaj}, it is not
difficult to convince oneself that the tunneling rate
from the symmetric
to the broken phase at $T<T_{c}$, which is
given in a volume $V$ by
$\Gamma\simeq V T_{c}^4{\rm exp}\left(-S_{3}/T\right)$, where $S_{3}$ is
the bounce action,  is extremely slow. Indeed, along the
ray $\p=R{\rm cos}\,\theta$ and $S=R{\rm sin}\,\theta$, where ${\rm
tan}\theta=\left(\bar{f}/v\right)$,  one obtains the naive
estimate
\beq
\frac{S_{3}}{T}=\frac{44\left(a_{\p}^2{\rm cos}^2\theta+a_{S}^2{\rm
sin}^2\theta\right)^{3/2}\left(4\pi\right)^2}{\left[G_{1}{\rm cos}^3\theta
+\left(1+3^{3/2}\right)\b^{3/2}{\rm sin}^3\theta\right]^2}\gg 1,
\eeq
where $G_{1}\equiv 6\left(g^2/2\right)^{3/2}+3\left[
\left(g^2+{g^{\prime}}^2\right)/2\right]^{3/2}\simeq 1.52$. Thus neither
the electroweak nor the lepton number transitions are expected to occur
at $T<T_{c}$. Nevertheless, as soon as the Universe cools down to a
temperature below $T_{QCD}\simeq 150$ MeV, because of the dynamical
breaking of the chiral symmetry and the appearance of a $\bar{\psi}\psi$
condensate, a new contribution to the finite
temperature potential becomes effective
\beq
\Delta V_{T}=-\v \zp^{3/2},
\eeq
where \cite{ga}
\beq
\v=g_{t}\langle\bar{\psi}\psi\rangle_{0}\left[1-\frac{N^2-1}{N}\frac{
T^2}{12 f_{\pi}^2}+{\cal O}\left(\frac{T^4}{\left(12\right)^2 f_{\pi}^4}
\right)\right],
\eeq
and $\langle\bar{\psi}\psi\rangle_{0}\simeq
-\left(250\,\mbox{MeV}\right)^{3}$, $N=6$, $f_{\pi}=93$ MeV. At
temperatures below $T_{QCD}$, a non-zero vacuum expectation value in the
direction $\zs=0$ is induced at
\beq
\a\simeq \frac{\v}{2a_{\p}}\frac{1}{T^2}.
\eeq
The false vacuum $\left(\zp=0,\zs=0\right)$ will roll down towards the
new unstable vacuum $\left(\zp=\a,\zs=0\right)$ very quickly, in a time
(in Hubble time unit)
\beq
\left(\frac{\Delta t}{H^{-1}}\right)\simeq
\frac{3}{2}\frac{g_{*}\left(T_{QCD}\right)}{a_{\p}}\left(\frac{T_{QCD}}{M_{P}}
\right)^2\ll 1,
\eeq
where $g_{*}\left(T_{QCD}\right)$ counts the effective degrees of
freedom at the temperature $T_{QCD}$.

When the temperature $T=T_{*}$, defined by\footnote{Note that
$\bar{\zp}^{1/2}\left(T_{*}\right)/T_{*}\simeq
\left(4\pi\right) a_{\p}/G_{1}\simless 1/g^2$, so that the high temperature
expansion is still valid.}
\beq
\varepsilon\left(T_{*}\right)=\frac{8\pi a_{\p}^2}{G_{1}}T_{*}^3,
\eeq
is reached, the metastability of the wrong unstable minimum $\a$ is lost
and the latter is free to roll down along the $\zs=0$
direction\footnote{Since the transition temperature is of order of
$\Lambda_{QCD}$ one may wonder about the size of perturbative and
nonperturbative QCD corrections to our effective Coleman-Weinberg-type
potential (8). Indeed, Flores and Sher \cite{sher} have argued that the
transition to the global minimum in the SM with a Coleman-Weinberg-type
potential does not take place since the
coupling of the quartic term in the potential for the Higgs field
changes sign for small values of $\p$. However in the SM, as well as
in the SMM, coupled to a JBD dilaton, the Higgs masses are large as
required by the positivity of the dilaton mass.}.

Expanding now the potential about the value $\left(\p_{+},0\right)$, we
get
\beq
V_{T}\left(\zs\right)=\left(a_{S} T^2+\g\p_{+}^2\right)\zs + \mbox{
other terms}.
\eeq
If $\g<0$ (looking at the Renormalization Group Equation for $\g$ one
can infer that $\g=0$ is a fixed point so that $\g$ does not change its
sign with the scale $M$), from eq. (46) we read off that the barrier
along the $\zs$ direction ceases to exist, and the lepton number is
spontaneously broken, when $\p_{+}$ reaches the value
$\left(-a_{S}/\g\right)^{1/2}T$, which is likely to occur at $T<T_{*}$
since
\beq
\left(a_{S} T_{*}^2+\g\bar{\zp}\left(T_{*}\right)\right)=
\as+\g\left(16\pi^2/G_{1}^2\right)a_{\p}^2>0
\eeq
for reasonable values of the parameters. Thus, if $\g<0$, it is the
electroweak phase transition to drive the breaking of $U(1)_{L}$.

If $\g>0$, from eq. (46) one could conclude that a barrier
in the $\zs$ direction is always present. This is not the case. Indeed,
eq. (46) is only valid until the expression $\langle\zp\rangle_{T}=
T^2/12$ may be used {\it i.e.} until $\p_{+}\simless T$. As soon as
$\p_{+}\simgreat T$, eq. (46) must be replaced by the more correct
expression
\beq
V_{T}=\left[\left(\frac{\sum_{i=1}^3\gi^2}{24}+
\frac{\ms}{12}k_{S}\right)T^2+\left(\g+\ms k_{\p}\right)\phi_{+}^2
\right]\zs + \mbox{other terms}.
\eeq
The presence of right-handed neutrinos is essential to keep the barrier
in the $z_{S}$ direction when $\p_{+}$ get values larger than $T$, {\it
i.e} when scalar fields no longer contribute to the $\sim T^2 z_{S}$
term. Assuming
that $\l\sim\b$ and $\m\sim\ms$, we conclude that, analogously to the case
$\g<0$, the $U(1)_{L}$
phase transition occurs after the electroweak transition, when $\p_{+}$
is of the order of $\sqrt{\left(\sum_{i=1}^3\gi^2-\l\right)/\left[24
\left(\l-\g\right)\right]}T$.

The change in the vacuum energy density as soon as the lepton number is
spontaneously broken is given by
\beq
\Delta V\simeq
a^4{\rm e}^{\frac{4\langle\s\rangle}{f}}\simeq
a^4\frac{\left(\l\langle\zp\rangle+\g\langle\zs\rangle\right)^2}{
m_{\p}^4}\simeq
\l T^{4}_{QCD}.
\eeq
Thus the maximum reheating of the radiation energy density that could
occur at this stage of the phase transitions is only ${\cal
O}\left(T_{QCD}^4\right)$, as it happens in the SM \cite{Mc}. Most of
the energy density of order $10^{-2}v^{4}$ will remain in the vacuum.

Quite surprisingly, solving the
FRW equations for the dilaton and Higgs field system one can
show that in the SM case most of the electroweak
phase transition goes into oscillations of the dilaton field after its
long slow roll along the valley floor down to its $T=0$ minimum at
$\s_{0}$. The point is that quantum fluctuations around $\s_{0}$ get
very small  masses, of order $v^2/f$, and they can decay only
into pairs of very light neutrinos with a rate $\Gamma_{\s}\sim
m_{\s}^3/f^2$. The temperature at the end of the $\s$ decays is then
$T_{RH}\simeq\left(M_{P}\Gamma_{\s}\right)^{1/2}\simeq\left(
M_{P}m_{\s}^3/f^2\right)^{1/2}$. To avoid the
Universe to be dominated by the energy stored in the
dilaton field during the nucleosynthesis and causing the calculation of
the light element abundances at present (which requires a radiation
dominated Universe \cite{ko} to be successful) to disagree with their
observed values, one must impose the bound $T_{RH}\simgreat 0.1$
MeV or, equivalently, the upper limit mentioned in the
Introduction, $f\simless 10^7$ GeV.

In the model discussed in this paper the situation is completely
different. Indeed, dilaton oscillations around the $T=0$ vacuum can
decay very efficiently into a pair of Majorons, the massless Goldstone
bosons associated to the breaking of the global $U(1)_{L}$ group.

Using the cartesian decomposition of the gauge singlet field
\beq
S=\rho+\bar{f}+iJ,
\eeq
where $J$ is the Majoron, and writing
\beq
{\rm e}^{2\s/f}= {\rm
e}^{2\langle\s\rangle/f}\left(1+\frac{2\s}{f}\right),
\eeq
from eq. (2) we read off the $\s JJ$ interaction term
\beq
{\cal L}_{int}=\frac{2\ms}{f} {\rm
e}^{2\langle\s\rangle/f}~\s JJ\simeq -2\frac{\b\bar{f}^2+\g v^2}{f}
{}~\s JJ.
\eeq
If had we used the more common decomposition of $S$ into polar
coordinates, $S=\left(\rho+\bar{f}\right){\rm
exp}\left(iJ/\bar{f}\right)$ the same interaction Lagrangian would have
been inferred from the Majoron kinetic term
\beq
{\cal L}_{kin}=\left[1+\left(\rho/\bar{f}\right)\right]^2
\partial_{\mu}J\partial^{\mu} J,
\eeq
which, after integrations by parts,
 yields the same interaction term of eq. (52)
\beq
\frac{1}{\bar{f}}\partial_{\mu}\partial^{\mu}\rho JJ=\frac{\ms}{f}
S{\rm
e}^{2\s/f}+....= \frac{2\ms}{f}{\rm
e}^{2\langle\s\rangle/f}~\s JJ+....
\eeq
The dilaton can decay into a pair of Majorons with a rate
\beq
\Gamma\left(\s\rightarrow JJ\right)=\frac{1}{8\pi}
\frac{m_{S}^4}{f^2m_{\s}}\simeq\frac{1}{8\pi}\b^2\frac{\bar{f}^2}{f}.
\eeq
Since $\Gamma\left(\s\rightarrow JJ\right)\gg H$, we conclude that the
Universe will be reheated
to a temperature\footnote{The pairs of Majorons produced by $\s$
decays rapidly thermalized with the thermal bath
through processes like $JJ\rightarrow \nu_{\tau}\nu_{\tau}$, with
$m_{\nu_{\tau}}\simeq$ 10 MeV \cite{st}.} ${\cal O}\left(10\right)$ GeV
as soon as the scalar fields relax to their $T=0$ minima after a long
slow roll down along the valley floor given by eq. (6). Thus, no upper
bound on $f$ has to be imposed since nucleosynthesis can now take place
in a standard way when the Universe is radiation-dominated.

The change in
entropy from the initial to the final state is $S_{f}/S_{i}=\left(
g_{* f}T_{f}^3/g_{* i}T_{i}^3\right)$, where $g_{* i\left(f\right)}$ is
the number of effectively massless particles in the finale (initial)
state, and $T_{f}$ $\left(T_{i}\right)$ are the final (initial)
temperature. With $T_{i}\sim T_{QCD}\sim$ 150 MeV, $T_{f}\sim$ 10 GeV,
$g_{* i}\sim 114.25$ and $g_{* f}\sim 87$ one obtains $S_{f}/S_i\sim
4\times 10^5$, which appears acceptable within the conventional
framework of baryon number generation in the early Universe
\cite{super}.

We finally point out that the presence of Majorons makes a heavy
$\tau$-neutrino, $m_{\nu_{\tau}}\simeq$ 10 MeV, cosmologically harmless
thanks to the fast decay $\nu_{\tau}\rightarrow \nu_{\mu,e}J$, in
contrast with the situation discussed in ref. \cite{neut}, where
neutrino masses were thought to be present at all scales through the
see-saw mechanism
and the rate of
the only possible decay $\nu_{\tau}\rightarrow \nu_{\mu,e}\s$ was not
large enough to avoid the decay products
to close the Universe, unless $f\simless 10^{9}$
GeV, again in contrast with the natural identification $f\simeq M_{P}$.
\vspace{0.5 truecm}
\begin{center}
{{\large {\bf 4. Conclusions}}}
\end{center}
\vspace{0.5 truecm}

The realization that any matter-antimatter asymmetry created at some
superheavy scale \cite{super} can be easily wiped out by $B$- and $L$-
SM quantum effects \cite{sph}, unless the baryon asymmetry is
proportional to the combination $\left(B-L\right)$, makes us faced with
the vital problem of avoiding that new $L$-violating interactions beyond
the SM are in thermal equilibrium when the SM anomalous effects are
still active. On the other hand, the SMM \cite{smm} naturally predicts
$L$-violating interactions and it has been recently shown in ref.
\cite{kari} that the global $U(1)_{L}$ group is spontaneously broken
before the electroweak phase transition so that a new mechanism to
regenerate the baryon asymmetry at the Fermi scale has to be invoked,
leading to a severe upper bound on the mass of the lightest neutral
Higgs from the requirement that the SM anomalous effects are
sufficiently suppressed after the accomplishment of the electroweak
phase transition \cite{bau}. Motivated by these considerations, we have
investigated an extension of the SMM with a dilaton field in which the
breaking of the scale invariance by the mass parameters of the scalar
potential is removed. Scale invariance remains broken by the dependence
of the couplings on the renormalization mass, {\it i.e} by the conformal
anomaly, and by the kinetic term of the dilaton. In curved space-time
this theory is precisely the Jordan-Brans-Dicke theory of gravity with
the SMM as matter sector \cite{Jor}.

Starting from the one-loop renormalization group improved potential we
have discussed the ground state of the theory. The classical potential
has a flat direction lifted by quantum corrections which have allowed us
to break the vacuum degeneracy and to find the true minima of the model.

We have then analyzed the cosmological consequences of a JBD dilaton on
the dynamics of the $SU(2)_{L}\otimes U(1)_{Y}$ and $U(1)_{L}$ phase
transitions in the early Universe. We have concluded that the lepton
number is spontaneously broken after the electroweak phase transition,
when anomalous effects are no longer active, which permits to escape any
strong bound on the couplings of the model. Furthermore, we have shown
that, contrary to the case of the SM coupled to a JBD dilaton, the scale
$f$ of the spontaneous breaking of scale invariance receives no limit
from considerations on the dilaton and/or neutrino cosmology, so that
we can still identify the scale $f$ with the Planck scale as required
to achieve a scalar-tensor theory of gravity and to provide an
explanation for the mass hierarchy between the Planck and the Fermi
scale \cite{hie}.
\vspace{0.3 cm}
\begin{center}
{{\large {\bf Acknowledgements}}}
\end{center}
\vspace{0.3 truecm}
It is a pleasure to express our gratitude to K. Enqvist, G.F. Giudice
and S. Matarrese for usuful and enlightening discussions and for reading
the early version of the paper.

\end{document}